\documentclass{article}

\usepackage{arxiv}

\usepackage[utf8]{inputenc} 
\usepackage[T1]{fontenc}    
\usepackage{url}            
\usepackage{booktabs}       
\usepackage{amsfonts}       
\usepackage{nicefrac}       
\usepackage{microtype}      
\usepackage{lipsum}         
\usepackage{graphicx}
\usepackage[numbers]{natbib}
\usepackage{float}
\usepackage[colorlinks=true,linkcolor=blue,citecolor=blue,urlcolor=blue]{hyperref}       
\usepackage{cleveref}       

\title{Unifying Speech Recognition, Synthesis and Conversion with Autoregressive Transformers}

\date{}

\newif\ifuniqueAffiliation
\uniqueAffiliationtrue

\ifuniqueAffiliation 
\author{ Runyuan Cai \quad Yu Lin \quad Yiming Wang \quad Chunlin Fu \quad Xiaodong Zeng\\
	\vspace{0.5em} \\
	\textit{AutoArk-AI} \\
	\vspace{0.5em} \\
	\texttt{\{runyuan.cai, yu.lin, yiming.wang, chunlin.fu, xiaodong.zeng\}@autoark.ai} \\
	\url{https://github.com/AutoArk/GPA}
}
\else
\usepackage{authblk}

\newbox{\orcid}\sbox{\orcid}{\includegraphics[scale=0.06]{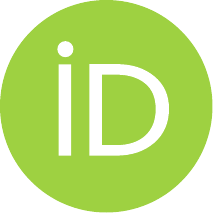}} 
\author[1]{Runyuan Cai}
\author[1]{Yu Lin}
\author[1]{Yiming Wang}
\author[1]{Xiaodong Zeng}
\author[1]{Chunlin Fu}
\affil[1]{AutoArk-AI}
\fi

\renewcommand{\undertitle}{Technical Report}
\renewcommand{\shorttitle}{Unifying Speech Recognition, Synthesis and Conversion with Autoregressive Transformers}

\hypersetup{
pdftitle={Unifying Speech Recognition, Synthesis and Conversion with Autoregressive Transformers},
pdfsubject={cs.SD, cs.CL, cs.AI},
pdfauthor={Runyuan Cai, Yu Lin, Yiming Wang, Xiaodong Zeng, Chunlin Fu},
pdfkeywords={Audio Generation, Speech Recognition, Voice Conversion, Foundation Model},
}

\begin{document}
\maketitle

\begin{abstract}
Traditional speech systems typically rely on separate, task-specific models for
text-to-speech (TTS), automatic speech recognition (ASR), and voice conversion (VC),
resulting in fragmented pipelines that limit scalability, efficiency, and cross-task
generalization.

In this paper, we present \textbf{General-Purpose Audio (GPA)}
\renewcommand{\thefootnote}{\fnsymbol{footnote}}
\footnote{Like an academic GPA that reflects capability across subjects, GPA aims for decent results across all audio tasks.}
\renewcommand{\thefootnote}{\arabic{footnote}},
a unified audio foundation model that integrates multiple core speech tasks within a
single large language model (LLM) architecture.
GPA operates on a shared discrete audio token space and supports instruction-driven task
induction, enabling a single autoregressive model to flexibly perform TTS, ASR, and VC
without architectural modifications.

This unified design combines a fully autoregressive formulation over discrete speech
tokens, joint multi-task training across speech domains, and a scalable inference
pipeline that achieves high concurrency and throughput.
The resulting model family supports efficient multi-scale deployment, including a
lightweight 0.3B-parameter variant optimized for edge and resource-constrained
environments.
Together, these design choices demonstrate that a unified autoregressive architecture
can achieve competitive performance across diverse speech tasks while remaining viable
for low-latency, practical deployment.
\end{abstract}


\section{Introduction}

The rapid advancement of generative artificial intelligence has fundamentally reshaped the landscape of speech processing. 
Driven by the success of Large Language Models (LLMs) in natural language processing, 
the speech community has moved from specialized, task-specific pipelines to large-scale, 
data-driven foundation models. Significant breakthroughs have been achieved in core tasks such as Text-to-Speech (TTS), 
Automatic Speech Recognition (ASR), and Voice Conversion (VC), 
where neural approaches now demonstrate human-level performance in distinct scenarios \cite{TTS1, ASR1, VC3}.

Despite these individual successes, the current ecosystem remains architecturally fragmented. 
While TTS, ASR, and VC share intrinsic linguistic and acoustic foundations, they are
typically treated as isolated problems, each relying on distinct modeling paradigms
and architectural assumptions:

\begin{itemize}
  \item \textbf{Text-to-Speech:}
  Modern zero-shot TTS systems have evolved into diverse categories, ranging from
  discrete token-based language models that autoregressively generate acoustic tokens
  \cite{TTS8,TTS9,TTS10,Spark}, to diffusion-based
  probabilistic mappings that implicitly learn alignments between text and speech
  \cite{TTS18,TTS26}, as well as coarse-to-fine
  hybrid architectures that combine semantic modeling with non-autoregressive acoustic
  rendering \cite{TTS31,TTS32,TTS-cozy}. While hybrid models currently
  offer a favorable balance between synthesis quality and controllability, their
  multi-stage design introduces significant system complexity, hindering unified
  modeling and efficient deployment.

  \item \textbf{Automatic Speech Recognition:}
  ASR has rapidly transitioned toward large-scale, pre-trained encoders and
  encoder--decoder models, such as Whisper \cite{ASR1}, Qwen-Audio \cite{ASR2,ASR3},
  SenseVoice \cite{ASR4}, and Seed-ASR \cite{ASR5}, demonstrating strong multilingual and
  multitask capabilities. However, these models are primarily optimized for
  discriminative recognition objectives and are architecturally specialized for ASR,
  making them difficult to repurpose for generative or conversion-oriented speech tasks
  without substantial architectural modification or auxiliary components
  \cite{ASR10,ASR11,ASR12,ASR13,ASR-firered}.

  \item \textbf{Voice Conversion:}
  VC systems generally rely on disentanglement-based frameworks that factorize speech
  into linguistic content, timbre, and speaking style to enable controllable attribute
  manipulation \cite{VC14,VC15}. Existing approaches include parallel-data-based
  conversion \cite{VC9,VC21}, disentangled representation learning
  \cite{VC23,VC24}, and large-scale in-context learning
  \cite{VC26,VC27,VC28}. Despite these advances, VC methods often suffer from
  limited scalability and imperfect decoupling of attributes, frequently requiring
  parallel corpora, explicit style annotations, or additional fine-tuning stages to
  achieve robust zero-shot generalization \cite{VC29,VC30,VC31,VC-VEVO}.
\end{itemize}

This fragmentation results in duplicated modeling efforts, limited cross-task knowledge transfer, 
and significant deployment complexity. 
To date, there is a lack of a \textbf{fully unified generative framework} that can natively support speech understanding,
generation, and conversion \textbf{within a single autoregressive modeling and inference pipeline},
without resorting to task-specific heads or routing mechanisms.

In this work, we present General-Purpose Audio (GPA), a unified speech foundation model that 
integrates TTS, ASR, and VC into a single autoregressive LLM framework. 
Unlike multi-stage or hybrid systems, GPA formulates all speech tasks as sequence modeling problems 
over a shared discrete audio token space. 
By conditioning on task-specific instructions, a single model can seamlessly switch 
between recognizing speech, synthesizing audio, and converting voices without architectural modification.

\noindent The main contributions of this work are summarized as follows:

\begin{itemize}
    \item \textbf{Unified Autoregressive Audio Framework:} 
	We propose GPA, a unified autoregressive LLM framework for TTS, ASR, and VC. 
	GPA employs a dual-tokenizer scheme, leveraging both GLM tokenizer\cite{GLM} and BiCodec tokenizer\cite{Spark} for speech discretization.
	By placing all tasks within a shared discrete audio token space, GPA enables instruction-driven task switching without any architectural modifications.

    \item \textbf{Synergistic Multi-Task Learning:} 
	We show that joint training across multiple tasks—using the Emilia \cite{Emilia} dataset 
	and supplementary proprietary data—enhances overall model performance. 
	This unified approach allows ASR, TTS, and VC to benefit from shared representations 
	in a discrete latent space, outperforming training each task in islation.
    
    \item \textbf{High-Throughput and Streaming Efficiency:} 
	We demonstrate that purely autoregressive architecture of GPA enables straightforward deployment, 
	superior concurrency, and higher overall throughput compared to hybrid alternatives. 
	This design choice makes GPA well-suited for streaming applications 
	and high-demand inference scenarios.
    
    \item \textbf{Edge-Optimized Scalability:}
	We introduce a family of GPA models spanning multiple scales, 
	including a highly compact 0.3B-parameter variant 
	explicitly designed for deployment in resource-constrained and edge environments. 
	This enables multi-task on-device inference within a single model instance 
	while maintaining a minimal memory footprint.

\end{itemize}

\section{General Purpose Audio}
Speech understanding, generation, and editing tasks such as ASR, TTS, and VC 
are commonly treated as distinct problems and addressed using task-specific pipelines. 
While effective, these designs fragment modeling choices and often rely on heterogeneous components. 

Although recent advances in LLMs have driven progress across speech tasks, hybrid approaches\cite{TTS-cozy} 
that depend on downstream synthesis modules\cite{fairytaler}\cite{scalablediffusion} remain constrained by latency and system efficiency. 
In contrast, the BiCodec system\cite{Spark} provides empirical evidence that 
a purely autoregressive architecture can achieve strong performance in TTS task. 

This success suggests that when speech and text are represented 
in a shared discrete token space, diverse audio tasks can be uniformly 
formulated as sequence-to-sequence generation problems, differing only 
in their input–output specifications.

Based on this insight, we arrive at a unifying principle: 
\textbf{all speech tasks admit a single autoregressive formulation in a fully discrete token space.}

\begin{figure}[!ht]
	\centering
	\includegraphics[width=0.8\textwidth]{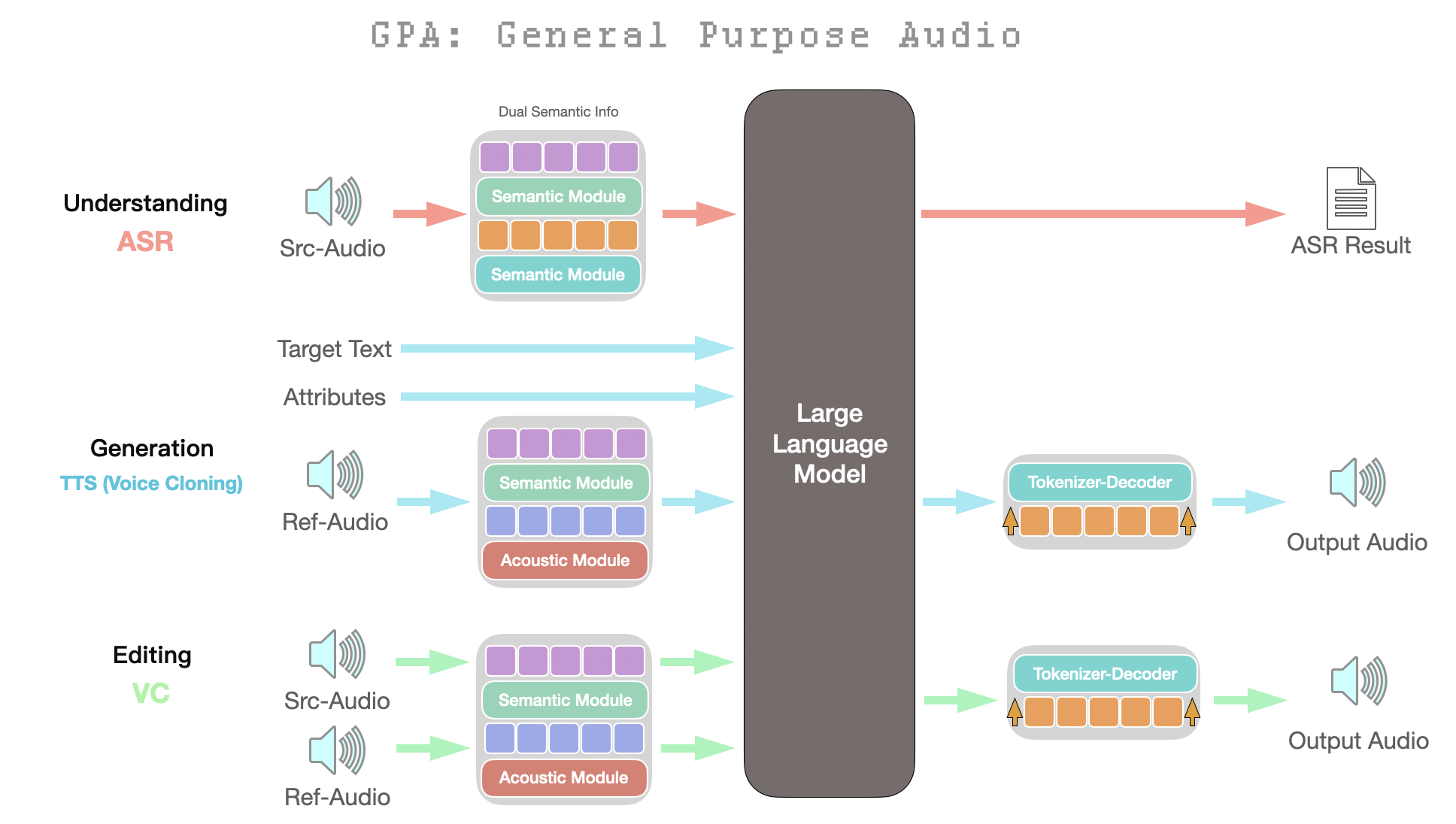}
	\caption{\textbf{Architecture of the proposed GPA framework.} The model utilizes a shared LLM backbone to unify three core audio tasks: understanding, generation, and editing. 
	Depending on the task, the model processes different combinations of inputs via Semantic and Acoustic modules 
	to generate the corresponding text or audio output.}
	\label{fig:gpa_architecture}
\end{figure}

Figure \ref{fig:gpa_architecture} presents the architecture of our proposed \emph{GPA} framework.
Building on Qwen3\cite{qwen3} LLM backbone, 
GPA adopts the BiCodec framework and 
integrates an extra GLM\cite{GLM} tokenizer to enhance semantic abstraction. 
By unifying textual, semantic, and acoustic information into a shared discrete token space, 
ASR, TTS, and VC can naturally emerge as instances of a single sequence generation task. 
This results in a general-purpose audio framework
based solely on autoregressive generation, with task behavior controlled entirely by instructions—
no task-specific architectures or decoding pipelines required.

Consequently, GPA eliminates the efficiency bottlenecks inherent in hybrid approaches, 
enabling high concurrency and throughput. 
The choice of the BiCodec representation paired with the GLM tokenizer 
further enhances the model's performance by promoting cross-task synergy.
We also demonstrate that joint training across multiple tasks 
within this unified framework yields mutual benefits, 
consistently outperforming single-task baselines.

\subsection{Motivation for a Fully Autoregressive Architecture}
We adopt a fully autoregressive architecture as the foundation of our model, 
unifying all audio tasks under a single generative formulation.

Speech tasks are inherently coupled in their underlying information space, 
sharing substantial representational overlap that is often underexploited 
by task-specific or modular designs. 
To fully leverage this shared structure, 
a unified framework capable of end-to-end joint training, as well as cross-task inference, is essential.

This requirement aligns with recent findings that when unimodal backbones are trained end-to-end, 
purely autoregressive models can outperform cross-attention architectures, 
despite having fewer parameters\cite{whatmatters}.
Such advantages are commonly attributed to the symmetric and homogeneous optimization of autoregressive models, 
whereas cross-attention introduces architectural asymmetry and gradient bottlenecks during joint training, 
limiting scalability in unified multimodal learning.

Consequently, a fully autoregressive formulation provides a more scalable and extensible foundation 
for joint training across heterogeneous audio tasks, 
while avoiding the structural constraints imposed by cross-attention-based designs.

\subsection{Tokenization and Codec Design}

BiCodec\cite{Spark} represents speech using complementary acoustic and semantic token streams, 
where the semantic tokens capture high-level information intrinsic to speech signals.
While effective, such semantic representations\cite{wav2vec2} are not explicitly aligned with 
discrete linguistic units present in human language.

Recent work\cite{SPHINX} 
has shown that combining pretrained representations from models 
with distinct architectures and pre-training paradigms enables richer and more robust semantic modeling. Motivated by this principle, we integrate the GLM~\cite{GLM} speech tokenizer—
derived from a large ASR model\cite{whisper-large-v3} under text-supervised training—
to enrich the unified token space with transcription-aligned semantics. 
Rather than replacing BiCodec's intrinsic semantic stream, 
GLM tokenizer complements it by providing linguistic units 
that correspond more closely to lexical and syntactic structures, 
which are inherently ambiguous or absent from unsupervised speech representations alone. 

Given the strong modeling capacity of modern LLM backbones, 
system performance is primarily limited by the semantic fidelity and linguistic grounding 
of the token representations. 
Therefore, despite the increased representational complexity 
introduced by these ASR-derived semantic tokens, 
we are able to supply the backbone with more linguistically coherent signals, 
leading to more expressive and controllable audio modeling without increasing model size.

\begin{figure}[!ht]
	\centering
	\includegraphics[width=0.8\textwidth]{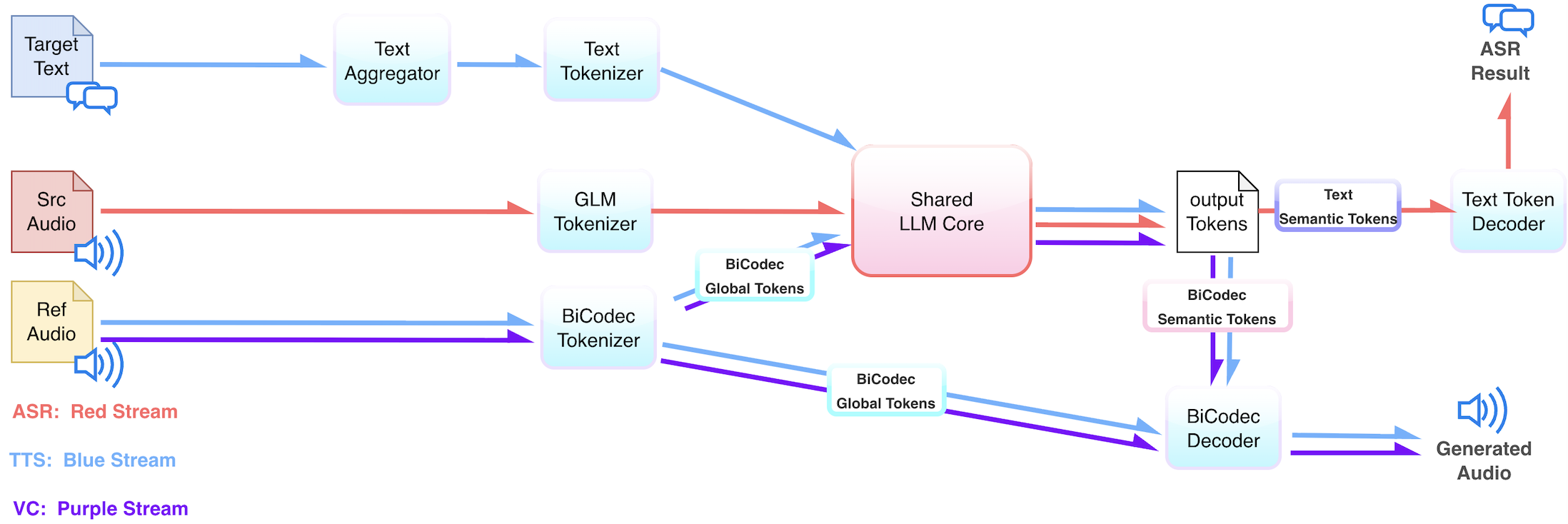}
	\caption{\textbf{Tokenization and task flow in GPA.} Different audio tasks are realized by varying the input and output token compositions, 
	all processed through a shared autoregressive backbone.}
	\label{fig:tokenization_flow}
\end{figure}

Under such design, speech understanding and production tasks—
including ASR, TTS, and voice conversion—
can leverage the same discrete semantic space. 
Figure~\ref{fig:tokenization_flow} details the tokenization and detokenization pathways 
each task follows while sharing the autoregressive backbone.

\subsection{Joint Multi-Task Training}
We train GPA in a joint multi-task manner using a mixture of the Emilia\cite{Emilia} dataset and a curated in-house corpus, 
covering ASR, TTS, and VC within a unified training schedule. 

Compared to task-specific training, joint optimization consistently improves performance across all tasks.
This improvement stems from the complementary inductive signals provided by different speech tasks 
when expressed in a shared discrete token space, 
a phenomenon also observed in prior joint modeling of speech generation objectives\cite{jointtrainingframework}. 
ASR enforces strong alignment between acoustic observations and linguistically grounded semantic tokens, 
anchoring the model's internal representations to stable lexical and syntactic structures. 
In contrast, TTS and VC emphasize faithful acoustic realization and speaker-dependent variability, 
encouraging the model to preserve fine-grained prosodic and timbral information conditioned on semantic inputs. 
Joint training therefore induces a balanced representation that is simultaneously linguistically grounded 
and acoustically expressive.

From an optimization perspective, 
the fully autoregressive formulation channels gradients 
from heterogeneous tasks through a shared next-token prediction objective, 
implicitly regularizing the backbone via diverse conditional generation patterns, 
consistent with observations from large autoregressive language models 
trained under a unified self-supervised objective \cite{radford2019language}.
Rather than competing for capacity, different tasks act as mutual constraints on the shared model, 
mitigating overfitting to task-specific biases and improving generalization. 

Related observations have been made in prior work 
that jointly trains speech recognition and synthesis objectives, 
suggesting that cross-task supervision can act as an implicit regularizer\cite{STTATTS}.
This effect is particularly pronounced when training on heterogeneous data sources\cite{SpeechT5}, 
where joint objectives encourage the model to learn task-invariant semantic abstractions 
while remaining robust to variations in speaker identity, recording conditions, and annotation quality.

Importantly, the unified tokenization scheme enables seamless task mixing without architectural modification 
or auxiliary task heads. 
All tasks are reduced to next-token prediction under different input–output compositions, 
allowing the model to benefit from increased data diversity and training signal density 
without increasing model size or inference complexity. 
As a result, joint multi-task training not only improves individual task performance, 
but also yields a more coherent and transferable audio foundation model.

\section{Empirical Evaluation}
To evaluate the GPA framework, 
we conduct empirical studies on its training dynamics and performance across diverse speech tasks.
We first describe the training protocol, 
including data composition, sample construction strategies, and fine-tuning procedures. 
Following this, we present quantitative and qualitative analyses of model performance, 
covering inference characteristics, latency, throughput, and task-specific evaluation metrics, 
to provide a thorough understanding of GPA's effectiveness and efficiency in practical scenarios.

\subsection{Training details}
This subsection describes the training protocol, data composition, 
and sample construction strategy used to train GPA, 
with an emphasis on scalability, reproducibility, 
and compatibility with unified autoregressive learning. 
To facilitate reproducibility and further research, 
the complete training code and configuration files are available 
at our public GitHub repository.\footnote{\url{https://github.com/AutoArk/GPA}}

\paragraph{From-Scratch Training.}
All GPA models are trained from scratch, 
without relying on pretrained speech or text generation weights. 
This design choice ensures that the learning process 
is free from external inductive biases introduced 
by language models, acoustic encoders, or task-specific architectures.

Instead, all model components are jointly optimized 
under a single next-token prediction objective. Consequently, 
differences in behavior across tasks—such as 
transcription, synthesis, or spoken intent understanding—
emerge solely from variations in the input–output token sequences, 
rather than from dedicated modules or separate prediction heads.

\paragraph{Pretraining Data.}
We conduct large-scale pretraining on approximately \textbf{1M} hours of speech data, 
drawn from a mixture of publicly available corpora, including Emilia~\cite{Emilia}, 
and a curated supplementary dataset constructed to enhance coverage of speaker diversity, 
recording conditions, and linguistic variability. 
Rather than targeting any single downstream task, 
this stage exposes the model to broad acoustic–semantic correspondences, 
establishing a robust foundation for unified speech understanding and generation.

\paragraph{Supervised Fine-Tuning.}
Following pretraining, we perform supervised fine-tuning on roughly \textbf{200K} hours of data 
covering ASR, TTS, and voice conversion. 
During this stage, training samples are constructed by instantiating different task-specific 
input--output token compositions over the same underlying token space.
This enables the model to adapt to precise task requirements 
while preserving the shared representations learned during large-scale pretraining.

\paragraph{Data Construction and Representation.}
To support unified speech modeling, all audio-text pairs are processed 
through a purpose-built curation pipeline that prioritizes 
semantic coherence, speaker coverage, and acoustic robustness.

\begin{itemize}
    \item \textbf{Audio normalization and denoising:} 
	Following previous work~\cite{TTS-cozy}, we apply dynamic range compression and
	peak-based volume normalization. Specifically, the waveform is rescaled by its
	maximum absolute amplitude:
	\[
	\tilde{x} = 0.6 \cdot \frac{x}{\max |x|},
	\]
	where $x$ denotes the raw waveform. This normalization ensures consistent amplitude scaling 
	across utterances, mitigating distributional shifts due to recording-level volume variations 
	and improving training stability.

	In addition, an in-house noise suppression model—trained on diverse real-world interference 
	(e.g., traffic, reverberation, overlapping speech)—
	is applied to enhance signal clarity.

    \item \textbf{Speaker-aware segmentation:} 
    An integrated VAD and speaker diarization system 
	jointly detects speech regions and assigns them to consistent speaker identities. 
	While our implementation relies on an internally developed module, 
	we encourage the use of prior work (e.g., \cite{pyannote}) to obtain similar functionality.

    This enables extraction of clean, speaker-homogeneous utterances 
	while discarding segments contaminated by cross-talk or non-speech events. 
    As part of this procedure, utterances exhibiting energy profiles 
	suggestive of mid-word truncation at boundaries 
	are automatically filtered out.

    \item \textbf{High-confidence transcription:} 
	Text labels are generated using a hybrid ASR strategy that leverages consensus across
	multiple publicly available models, including Faster-Whisper Large-V3\cite{whisper-large-v3},
	Canary-1B \& Parakeet-TDT-0.6B-v3\cite{CanaryParakeet}, Omnilingual ASR\cite{omnilingual},
	and SeamlessM4T-v2-Large\cite{seamlessm4t}.
	Each model produces an independent transcription hypothesis, which are subsequently
	combined and scored by an in-house confidence-aware module.

	Only segments exhibiting strong cross-system agreement are retained.
	In practice, agreement is quantified using the average pairwise word error rate (pWER)
	across all ASR hypotheses:
	\[
	\mathrm{pWER} = \frac{1}{N(N-1)} \sum_{i \neq j} \mathrm{WER}(y_i, y_j),
	\]
	where $\{y_1, \ldots, y_N\}$ denote the transcription outputs of $N$ ASR models.
	Following the setup in previous work\cite{TTS-cozy}, segments with $\mathrm{pWER} < 15\%$ are retained as final text labels.

    \item \textbf{Prosody-aligned punctuation:} 
    Using forced alignment from Montreal Forced Aligner~\cite{mfa}, inter-word pause durations inform punctuation refinement: commas are inserted for silences $\ge$300 ms between clauses, while spurious punctuation in fluent spans (<50 ms pauses) is removed. 
    This yields a temporally grounded and linguistically coherent text stream, ready for joint discretization into a shared speech–text token space.
\end{itemize}

\subsection{Model Performance}
The effectiveness of a unified speech model in real-world deployment depends not only
on its accuracy, but also on the efficiency and scalability of its inference pipeline.
In this subsection, we analyze the performance characteristics of GPA from both an
architectural and a system-level perspective.

We first discuss the inference behavior of GPA's purely autoregressive design,
highlighting how its streaming-friendly formulation enables simple batching,
high concurrency, and efficient utilization of compute resources.
We then provide quantitative evidence of these properties through latency and
throughput measurements under representative TTS and ASR streaming workloads,
focusing on the deployment-oriented \textbf{GPA-0.3B} configuration optimized for low latency
and memory efficiency.

Together, these analyses illustrate that GPA is well suited for both large-scale
server deployment and latency-sensitive, resource-constrained execution scenarios.

\subsubsection{Inference}
GPA adopts a purely autoregressive inference paradigm, in which both text and speech
are represented as discrete tokens and generated sequentially by a single shared
Transformer backbone. All speech tasks—including ASR, TTS, and voice conversion—are
handled within the same decoding framework, without switching models or invoking
task-specific components at inference time. Task behavior is instead specified
implicitly through token ordering and input prompting, resulting in a unified and
lightweight inference pipeline.

This design naturally lends itself to streaming scenarios. Tokens are emitted
incrementally as decoding progresses, enabling low-latency output without requiring
explicit look-ahead mechanisms or multi-stage buffering. Compared to conventional
pipelines that rely on separate encoders, decoders, or post-processing stages, GPA
maintains a single, continuous generation process, reducing both system complexity
and coordination overhead.

From a deployment perspective, the simplicity of autoregressive decoding allows GPA
to integrate seamlessly with standard LLM inference frameworks. In practice, this
translates to favorable throughput and latency characteristics under concurrent
workloads, particularly in streaming settings. The same inference procedure applies
consistently across cloud and edge environments, making GPA well suited for scalable
real-time speech applications. 

To facilitate reproducibility and practical adoption,
we release the complete inference and deployment codebase as part of our public release.

\subsubsection{latency and throughput}
We report streaming latency and throughput for both TTS and ASR 
using the \textbf{GPA-0.3B} model under varying request concurrency.
For TTS, we measure time-to-first-chunk (TTFC) and real-time factor (RTF), 
where lower values indicate faster first audio delivery and higher synthesis speed, respectively. 
For ASR, we report time-to-first-token
(TTFT) and end-to-end latency. 

All results are measured in a streaming setting, highlighting how system
responsiveness and overall throughput evolve as concurrent load increases.
Reproduction can be obtained by following the official deployment scripts\footnote{\url{https://github.com/AutoArk/GPA}}, 
which reflects end-to-end performance in realistic serving scenarios rather than offline inference.

\begin{table}[t]
    \centering
    \resizebox{\textwidth}{!}{%
    \begin{tabular}{r r r r r r r r}
        \toprule
        \textbf{Concurrency} &
        \textbf{Avg TTFC (ms)} &
        \textbf{P50 TTFC (ms)} &
        \textbf{P99 TTFC (ms)} &
        \textbf{Avg RTF} &
        \textbf{P50 RTF} &
        \textbf{P99 RTF} &
        \textbf{Audio Dur (s)} \\
        \midrule
        1   & 258.8  & 258.8  & 258.8  & 0.197 & 0.197 & 0.197 & 6.44 \\
        5   & 385.0  & 394.7  & 396.2  & 0.218 & 0.217 & 0.248 & 6.76 \\
        10  & 544.6  & 564.2  & 566.7  & 0.282 & 0.301 & 0.313 & 6.49 \\
        20  & 977.8  & 977.9  & 982.9  & 0.470 & 0.490 & 0.538 & 7.19 \\
        40  & 1797.0 & 1736.4 & 2564.5 & 0.421 & 0.400 & 0.587 & 6.33 \\
        80  & 3786.4 & 4054.4 & 5415.8 & 0.763 & 0.763 & 1.096 & 6.32 \\
        160 & 9847.9 & 10239.9 & 14350.3 & 1.718 & 1.740 & 2.577 & 6.44 \\
        \bottomrule
    \end{tabular}%
    }
    \vspace{4pt}
    \caption{TTS streaming benchmark under varying concurrency levels.
    TTFC denotes time-to-first-chunk (ms), RTF is the real-time factor, and Audio Dur
    indicates the average generated audio duration in seconds.}
    \label{tab:tts_streaming_latency_throughput}
\end{table}

\begin{table}[t]
    \centering
    \begin{tabular}{r r r r r}
        \toprule
        \textbf{Concurrency} &
        \textbf{Avg TTFT (ms)} &
        \textbf{P50 TTFT (ms)} &
        \textbf{P99 TTFT (ms)} &
        \textbf{Avg Total (ms)} \\
        \midrule
        1   & 157.5  & 157.5  & 157.5  & 190.9 \\
        5   & 394.1  & 393.7  & 395.9  & 400.0 \\
        10  & 589.6  & 721.3  & 723.3  & 598.1 \\
        20  & 1316.3 & 1495.6 & 1500.4 & 1317.8 \\
        40  & 2690.9 & 2678.3 & 2861.4 & 2693.7 \\
        80  & 3833.4 & 3961.3 & 4027.0 & 3845.1 \\
        160 & 5037.0 & 5689.3 & 6676.0 & 5044.0 \\
        \bottomrule
    \end{tabular}
    \vspace{4pt}
    \caption{ASR streaming latency as a function of concurrency.
    TTFT denotes time-to-first-token. All values are reported in milliseconds.}
    \label{tab:asr_streaming_latency}
\end{table}

\subsection{Model Evaluations}
We evaluate GPA on representative TTS and ASR benchmarks to assess both its
practical effectiveness and its scalability across model sizes.
Experiments are conducted using two model variants with complementary design goals.
Voice conversion (VC) is not evaluated in a separate table, as it shares the same
automatic evaluation metrics as TTS, namely intelligibility-related error rates 
and speaker similarity.
Since VC in GPA differs from TTS primarily in conditioning and token composition
rather than in the underlying autoregressive generation mechanism, separate numerical
evaluation would be redundant.

The compact GPA-0.3B model is optimized for memory efficiency and on-device
deployment, targeting resource-constrained scenarios where low footprint and
streaming capability are primary considerations.
While it does not aim to match the strongest large-scale systems, it achieves
competitive performance within its parameter regime.
In contrast, the GPA-3B model serves as a capability-oriented configuration,
demonstrating the effectiveness of the proposed unified autoregressive framework
when model capacity is scaled up.

For text-to-speech evaluation(Table~\ref{tab:tts_eval}), we report Character Error Rate (CER) and speaker
similarity (Sim) on Chinese test sets, and Word Error Rate (WER) together with
speaker similarity on English benchmarks, following established evaluation
protocols~\cite{TTS26}.
Lower error rates indicate better intelligibility, while higher similarity scores
reflect improved speaker consistency.
For automatic speech recognition, we report standard WER or CER metrics on
Librispeech~\cite{LibriSpeech}, and
AISHELL-1~\cite{AISHELL}, covering both English and Mandarin speech recognition
settings. 

These benchmarks span both acoustic fidelity and linguistic competence across
tasks and languages.
ASR performance, in particular, depends on both acoustic modeling and the ability
to capture long-range linguistic structure.
As a result, recognition accuracy varies across model scales (Table~\ref{tab:asr_eval}),
reflecting different trade-offs between representational capacity and deployment
efficiency.
Together, these results characterize GPA's behavior across tasks, languages,
and model configurations.

\begin{table}[htbp]
    \centering
    \small
    \setlength{\tabcolsep}{3pt}
    \begin{tabular}{lcccccc}
        \toprule
        Model & Open-Source & Model Size & \multicolumn{2}{c}{Seed-zh\cite{TTS26}} & \multicolumn{2}{c}{Seed-en\cite{TTS26}} \\
        \cmidrule(lr){4-5} \cmidrule(lr){6-7}
         & & & CER (\%) $\downarrow$ & Sim (\%) $\uparrow$ & WER (\%) $\downarrow$ & Sim (\%) $\uparrow$ \\
        \midrule
        \multicolumn{7}{c}{\textbf{Multi-Stage or NAR Methods}} \\
        \midrule
        Human & - & - & 1.26 & 75.5 & 2.14 & 73.4 \\
        Seed-TTS \cite{TTS26} & No & - & 1.12 & \textbf{79.6} & 2.25 & \textbf{76.2} \\
        MiniMax-Speech \cite{MiniMax} & No & - & 0.83 & 78.3 & \textbf{1.65} & 69.2 \\
        F5-TTS \cite{fairytaler} & Yes & 0.3B & 1.52 & 74.1 & 2.00 & 64.7 \\
        CosyVoice2 \cite{cosy2} & Yes & 0.5B & 1.45 & 75.7 & 2.57 & 65.9 \\
        FireRedTTS2 \cite{fireredtts} & Yes & 1.5B & 1.14 & 73.2 & 1.95 & 66.5 \\
        Index-TTS2 \cite{IndexTTS2} & Yes & 1.5B & 1.03 & 76.5 & 2.23 & 70.6 \\
        VibeVoice-1.5B \cite{VibeVoice} & Yes & 1.5B & 1.16 & 74.4 & 3.04 & 68.9 \\
        VibeVoice-Realtime \cite{VibeVoice} & Yes & 0.5B & - & - & 2.05 & 63.3 \\
        HiggsAudio-v2 & Yes & 3B & 1.50 & 74.0 & 2.44 & 67.7 \\
        VoxCPM \cite{VoxCPM} & Yes & 0.5B & 0.93 & 77.2 & 1.85 & 72.9 \\
        GLM-TTS \cite{GLM} & Yes & 1.5B & 1.03 & 76.1 & - & - \\
        GLM-TTS RL \cite{GLM} & Yes & 1.5B & 0.89 & 76.4 & - & - \\
        Fun-CosyVoice3-0.5B-2512 \cite{TTS-cozy} & Yes & 0.5B & 1.21 & 78.0 & 2.24 & 71.8 \\
        Fun-CosyVoice3-0.5B-2512\_RL \cite{TTS-cozy} & Yes & 0.5B & \textbf{0.81} & 77.4 & 1.68 & 69.5 \\
        \midrule
        \multicolumn{7}{c}{\textbf{One-Stage AR Methods}} \\
        \midrule
        Spark TTS \cite{Spark} & Yes & 0.5B & 1.20 & 66.0 & 1.98 & 57.3 \\
        \textbf{GPA-0.3B} & Yes & 0.3B & \textbf{0.95} & 65.9 & \textbf{1.51} & 56.5 \\
        \textbf{GPA-3B} & Yes & 3B & \textbf{0.84} & 73.0 & \textbf{1.31} & 71.7 \\
        \bottomrule
    \end{tabular}
    \vspace{4pt}
    \caption{TTS Evaluation Table. 
    Results are grouped into multi-stage (NAR) and one-stage autoregressive (AR) methods. 
    We report Character Error Rate (CER) and Speaker Similarity (Sim) for Chinese (Seed-zh), 
    and Word Error Rate (WER) and Speaker Similarity (Sim) for English (Seed-en) 
    following the evaluation protocol in \cite{TTS-cozy}. 
    $\downarrow$ indicates lower is better and $\uparrow$ indicates higher is better.}
    \label{tab:tts_eval}
\end{table}

\begin{table}[htbp]
    \centering
    \small
    \begin{tabular}{lccc}
    \toprule
    Model & Params & Librispeech test-clean (WER↓) & AISHELL-1 (CER↓) \\
    \midrule

    \multicolumn{4}{l}{\textbf{Models with $<$ 0.5B parameters}} \\
    \midrule
    Whisper-S \cite{whisper-large-v3} & 0.24B & 3.13 & -- \\
    \textbf{GPA-0.3B} & 0.3B & 8.88 & 4.50 \\

    \midrule
    \multicolumn{4}{l}{\textbf{Models with $\ge$ 0.5B parameters}} \\
    \midrule
    Fun-ASR-nano\cite{funasr} & 0.8B & 1.76 & 1.80 \\
    FireRed-ASR\cite{ASR-firered} & 1.1B & 1.84 & 0.54 \\
    GLM-ASR-nano\cite{GLM} & 1.5B & 2.00 & 1.81 \\
    GLM-ASR-nano*\cite{GLM} & 1.5B & 2.17 & 2.17 \\
    Whisper-L\cite{whisper-large-v3} & 1.55B & 1.82 & 4.72 \\
    Kimi-Audio\cite{kimiaudio} & -- & 1.32 & 0.71 \\
    Step-Audio2\cite{stepaudio2} & -- & 1.17 & 0.63 \\
    Seed-ASR\cite{ASR5} & -- & 1.58 & 0.68 \\
    Seed-ASR*\cite{ASR5} & -- & 2.80 & 1.63 \\
    Fun-ASR\cite{funasr} & 7.7B & 1.51 & 1.22 \\
    \textbf{GPA-3B} & 3.0B & 2.52 & 1.93 \\

    \bottomrule
    \end{tabular}
    \vspace{4pt}
    \caption{
        ASR evaluation is conducted on Librispeech and AISHELL-1.
        WER (\%) is reported for Librispeech, while CER (\%) is reported for AISHELL-1.
        Given the strong scaling behavior of language modeling in ASR, results are grouped
        and compared by model size.
        Baseline results are taken from published reports\cite{GLM,omniflatten} under the same evaluation protocol.
        Models with undisclosed parameter counts are placed at the bottom of the table.
        Seed-ASR* results are obtained via the official Volcengine API, while GLM-ASR-nano*
        results are evaluated using the open-source checkpoint. 
        }
    \label{tab:asr_eval}
\end{table}

\clearpage

\section{Conclusion}
In this paper, we introduced GPA, a unified autoregressive framework for general-purpose
speech modeling that consolidates multiple speech tasks within a single LLM-based
architecture.
By representing speech, text, and control signals as discrete tokens and modeling
them under a shared next-token prediction objective, GPA supports automatic speech
recognition, text-to-speech synthesis, and speech editing within a common formulation.
This unified design reduces architectural fragmentation and enables efficient
training and inference across tasks.

Comprehensive evaluations show that GPA achieves competitive accuracy on a range of
speech benchmarks while maintaining strong efficiency and scalability properties.
These results validate the effectiveness of a fully autoregressive, token-based
approach to unified speech modeling and highlight its potential for building
flexible and deployable audio foundation models.

\section{Limitations and Future Work}
While GPA demonstrates the practicality of a unified autoregressive framework for
multiple speech tasks, several limitations remain.
First, although the shared token-based formulation simplifies system design, it
introduces a single modeling bottleneck, which may limit peak performance on
highly specialized tasks when compared to purpose-built architectures.
Second, as with other large autoregressive models, inference cost and latency scale
with sequence length, posing challenges for extremely long-form or low-latency
applications without further optimization.
Third, ASR performance of the GPA-0.3B model is comparatively weaker, primarily due to
capacity constraints rather than fundamental limitations of the unified autoregressive
design, indicating clear benefits from scaling and further optimization.

Looking ahead, the unified autoregressive formulation of GPA naturally admits
reinforcement learning over discrete audio tokens.
Unlike heterogeneous pipelines that combine separate acoustic models, language
models, and task-specific modules—where credit assignment across components is
often ambiguous—GPA's single next-token prediction objective provides a clear and
consistent optimization target.
This property opens the door to metric-driven post-training, enabling direct
optimization for instruction following, perceptual quality, and task-specific
objectives within a unified framework.

\clearpage
\bibliographystyle{plain}
\bibliography{references} 

\end{document}